\documentclass[prl,aps,onecolumn,superscriptaddress,showpacs,showkeys,preprint]{revtex4}

\usepackage{epsfig}
\usepackage{latexsym,amssymb,amsmath}

\newcommand{\eq}[1]{\begin{align} #1 \end{align}}

\begin{document}

\today

\title{Multiplicity Fluctuations in Au+Au Collisions at RHIC}

\author{V.P.~Konchakovski}
\affiliation{Bogolyubov Institute for Theoretical Physics, Kiev,
Ukraine} \affiliation{Helmholtz Research School, University of
Frankfurt, Frankfurt, Germany}
\author{M.I.~Gorenstein}
\affiliation{Bogolyubov Institute for Theoretical Physics, Kiev, Ukraine}
\affiliation{Frankfurt Institute for Advanced Studies, Frankfurt, Germany}
\author{E.L.~Bratkovskaya}
\affiliation{Frankfurt Institute for Advanced Studies, Frankfurt,
Germany}

\begin{abstract}
The preliminary data of the PHENIX collaboration for the scaled
variances of charged hadron multiplicity fluctuations in Au+Au at
$\sqrt{s}=200$ GeV are analyzed within the model of independent
sources.  We use the HSD transport model to calculate the participant
number fluctuations and the number of charged hadrons per nucleon
participant in different centrality bins. This combined picture leads
to a good agreement with the PHENIX data and suggests that the measured
multiplicity fluctuations result dominantly from participant number
fluctuations.  The role of centrality selection and acceptance is
discussed separately.
\end{abstract}



\maketitle


The event-by-event fluctuations in high energy nucleus-nucleus
(A+A) collisions (see e.g., the reviews \cite{rev1,rev2}) are
expected to provide signals of the transition between different
phases (see e.g., Refs.~\cite{ood,fluc2}) and the QCD critical
point \cite{fluc3}. In the present letter we study the charged
multiplicity fluctuations in Au+Au collisions at RHIC energies.
The preliminary data of the PHENIX collaboration \cite{PHENIX} at
$\sqrt{s}=200$~GeV are analyzed within the model of independent
sources while employing the microscopic Hadron-String-Dynamics~(HSD)
transport model \cite{HSD,Weber} to define the centrality selection
and to calculate the properties of hadron production
sources.

The centrality selection is an important aspect of fluctuation
studies in A+A collisions. At the SPS fixed target experiments the
samples of collisions with a fixed number of projectile
participants $N_P^{proj}$ can be selected to minimize the
participant number fluctuations in the sample of collision events.
This selection is possible due to a measurement of the number of
nucleon spectators from the projectile,  $N_S^{proj}$, in each
individual collision by a calorimeter which covers the projectile
fragmentation domain. However, even in the sample with $N_P^{proj}
= const$ the number of target participants fluctuates
considerably. In the following the variance, $Var(n) \equiv
\langle n^2 \rangle - \langle n \rangle^2$, and scaled variance,
$\omega \equiv Var(n)/\langle n \rangle$, where $n$ stands for a
given random variable and $\langle \cdots \rangle$ for
event-by-event averaging, will be used to quantify fluctuations.
In each sample with $N_P^{proj}=const$ the number of target
participants fluctuates around its mean value, $\langle N_P^{targ}
\rangle=N_P^{proj}$, with the scaled variance $\omega_P^{targ}$.
Within the HSD and UrQMD transport models it was found in Ref.
\cite{voka1} that the fluctuations of $N_ P^{targ}$ strongly
influence the charged hadron fluctuations. The constant values of
$N_P^{proj}$ and fluctuations of $N_P^{targ}$ lead also to an
asymmetry between the fluctuations in the projectile and target
hemispheres. The consequences of this asymmetry depend on the A+A
dynamics as discussed in Ref.~\cite{MGMG}.

\begin{figure*}[h!]
\epsfig{file=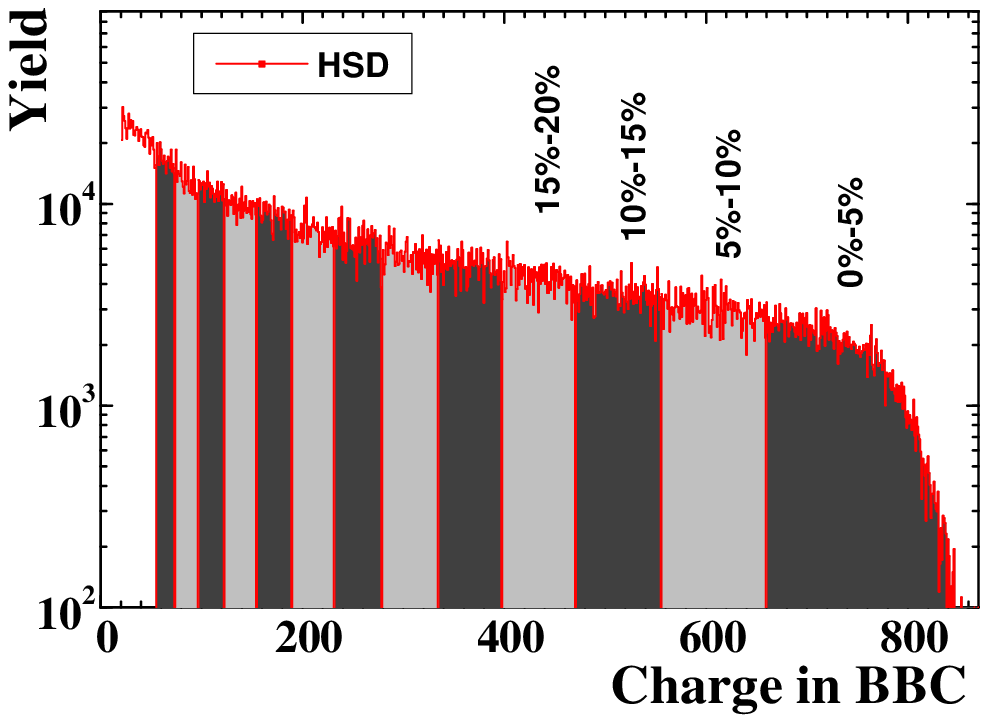,width=8cm}
\epsfig{file=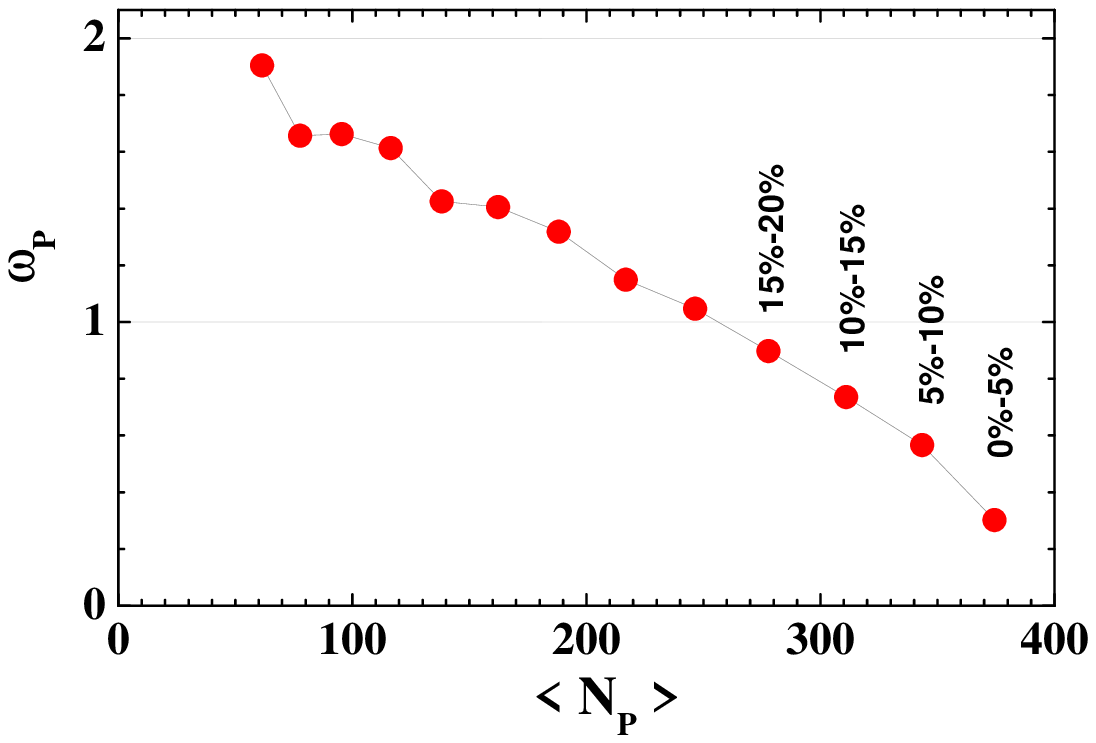,width=8.3cm}
 \caption{HSD model results for Au+Au collisions at $\sqrt{s}=200$~GeV.
 {\it Left:} Centrality classes
defined via the BBC distribution. {\it Right:} The average
number of participants, $\langle N_P\rangle$, and the scaled
variance  of the participant number fluctuations, $\omega_P$,
calculated for the 5\%  BBC centrality classes. } \label{BBC}
\end{figure*}

In Au+Au collisions at  RHIC a different centrality selection is
used. There are two kinds of detectors which define the centrality
of Au+Au collision, Beam-Beam Counters (BBC) and Zero Degree
Calorimeters (ZDC). At the c.m. pair energy
$\sqrt{s}=200$~GeV, the BBC measure the charged particle
multiplicity in the pseudorapidity range $3.0<|\eta|<3.9$, and the
ZDC -- the number of neutrons with $|\eta|>6.0$ \cite{PHENIX}.
These neutrons are part of the nucleon spectators. Due to
technical reasons the neutron spectators can be only detected by
the ZDC (not protons and nuclear fragments), but in both
hemispheres. The BBC distribution will be used in the HSD
calculations to divide Au+Au collision events into 5\% centrality
samples. HSD does not specify different spectator groups --
neutrons, protons, and nuclear fragments such that we can not use the
ZDC information. In Fig.~\ref{BBC} (left) the HSD results for the
BBC distribution and centrality classes in Au+Au collisions at
$\sqrt{s}=$200~GeV are shown. We find a good agreement between the
HSD shape of the BBC distribution and the PHENIX data
\cite{PHENIX}. The experimental estimates of $\langle N_P\rangle$
for different centrality classes are based on the Glauber model.
These estimates vary by less than 0.5\% depending on the shape of
the cut in the ZDC/BBC space or whether the BBC alone is used as a
centrality measure \cite{PHENIX}. Note, however, that the HSD
$\langle N_P\rangle$ numbers are not exactly equal to the PHENIX
values. It is also not obvious that different definitions for the
5\% centrality classes give the same values of the scaled variance
$\omega_P$ for the participant number fluctuations.

\begin{figure*}[h!]
\epsfig{file=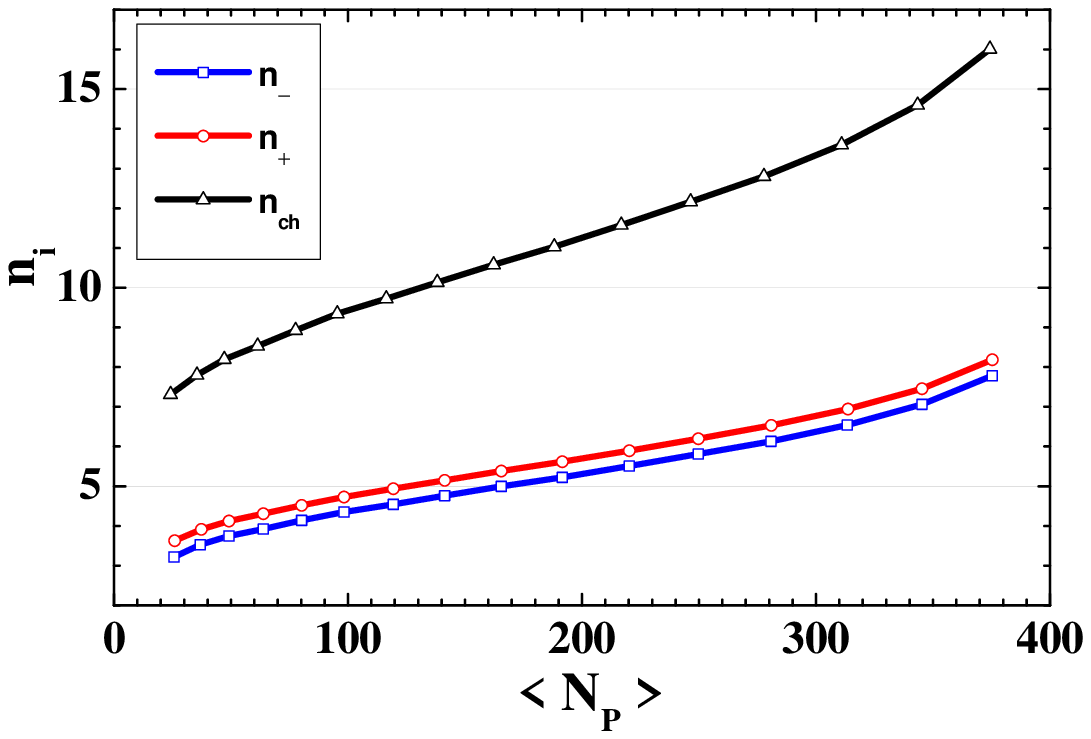,width=8cm}
\epsfig{file=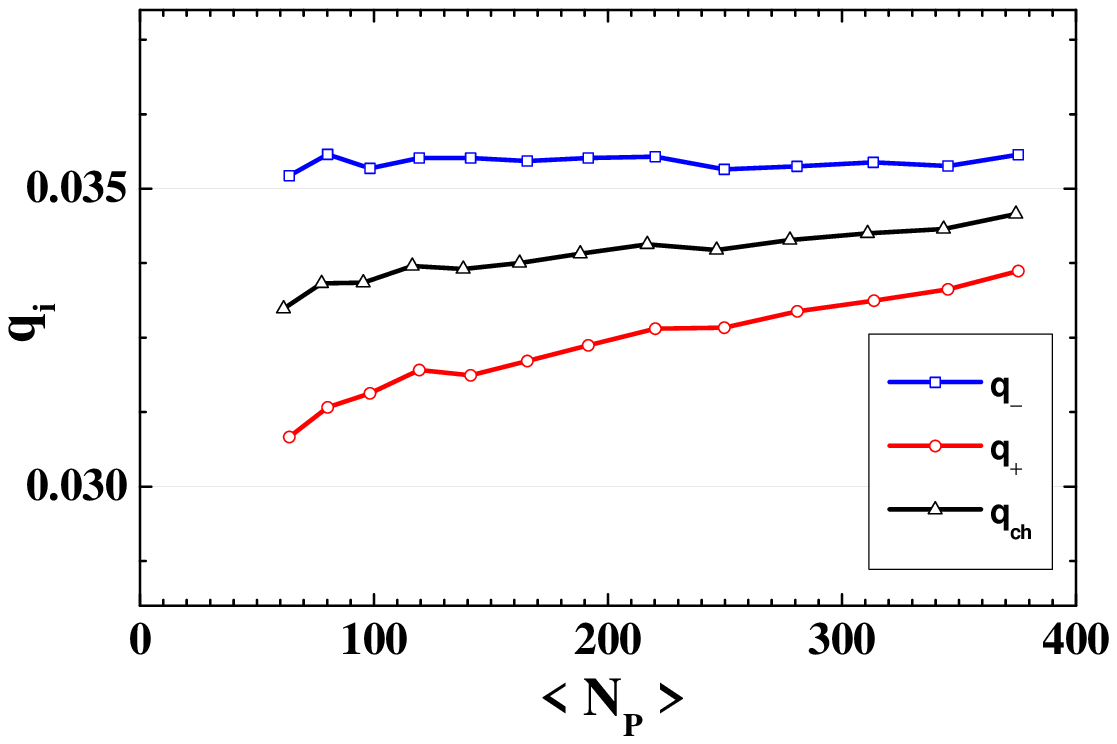,width=8cm}
 \caption{HSD results for different BBC centrality classes in
 Au+Au collisions at $\sqrt{s}=200$~GeV.
{\it Left:} The mean number of charged hadrons per participant,
$n_{i}~=~\langle N_{i}\rangle/\langle N_P\rangle$. {\it Right:}
The fraction of accepted particles, $q_i=\langle
N^{acc}_{i}\rangle/\langle N_{i}\rangle$.
 } \label{omTeor}
\end{figure*}

Defining the centrality selection via the HSD transport model (which is
similar to the BBC in the PHENIX experiment) we calculate the mean
number of nucleon participants, $\langle N_P\rangle$, and the
scaled variance of its fluctuations, $\omega_P$,  in each 5\%
centrality sample. The results are shown in Fig.~\ref{BBC}, right. The
Fig.~\ref{omTeor} (left) shows the HSD results for the mean number
of charged hadrons per nucleon participant, $n_{i}=\langle
N_{i}\rangle/\langle N_P\rangle$, where  the index $i$ stands for ``$-$'',
``+'', and ``ch'', i.e negatively, positively, and all charged
final hadrons. Note that the centrality dependence of $n_i$ is
opposite to that of $\omega_P$: $n_i$ increases with $\langle
N_P\rangle$, whereas $\omega_P$ decreases.

The PHENIX detector accepts charged particles in a small region of the
phase space with pseudorapidity $|\eta|<0.26$ and azimuthal angle
$\phi<245^o$ and the $p_T$ range from 0.2 to 2.0 GeV/c
\cite{PHENIX}. The fraction of the accepted particles
$q_i=\langle N^{acc}_{i}\rangle/\langle N_{i}\rangle$ calculated within
the HSD model is shown in the r.h.s. of Fig.~\ref{omTeor}.  According to the
HSD results  only  $3\div 3.5$\% of charged particles are
accepted by the mid-rapidity PHENIX detector.

\begin{figure*}[h!]
\vspace*{1mm}
\epsfig{file=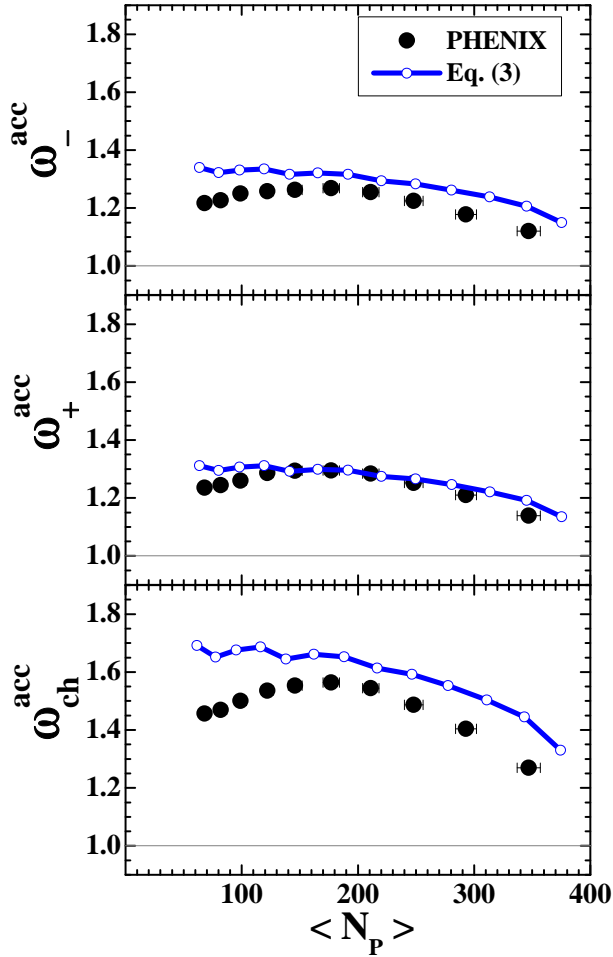,width=8cm}
\caption{The scaled variance of charged particle fluctuations in Au+Au
collisions at $\sqrt{s}=200$~GeV with the PHENIX acceptance. The circles
are the PHENIX data \cite{PHENIX} while the open points (connected
by the solid line) correspond to Eq.~(\ref{omega-acc}) with the
HSD results for $\omega_P$, $n_i$, and $q_i$. }
\label{comp}
\end{figure*}

To estimate the role of the participant number event-by-event
fluctuations we use the model of independent sources
(see~e.g., Refs~\cite{rev1,voka1,MGMG}),
%
\eq{\label{WMod} \omega_i~=~\omega^*_i~+~n_i~ \omega_P~. }
 The first term in the
r.h.s. of Eq.~(\ref{WMod}) corresponds to the fluctuations of
the hadron multiplicity from one source, and the second term,
$n_i~\omega_P$, gives additional fluctuations due to the
fluctuations of the number of sources. As usually, we have assumed
that the number of sources is proportional to the number of
nucleon participants. The value of $n_i$ in Eq.~(\ref{WMod}) is
then the average number of $i$'th particles per participant,
$n_i=\langle N_i\rangle /\langle N_P\rangle$. We also assume that
nucleon-nucleon collisions define the fluctuations $\omega_i^*$
from a single source. To calculate the fluctuations
$\omega^{acc}_i$ in the PHENIX acceptance we use the acceptance
scaling formula (see e.g., Refs.~\cite{rev1,voka1,MGMG}):
\eq{\label{accsc}
\omega^{acc}_i~=~1~-q_i~+~q_i~\omega_i~, }
where $q_i$ is the fraction of the accepted $i$'th hadrons by the
PHENIX detector. Using Eq.~(\ref{WMod}) for $\omega_i$ one finds,
\eq{
\omega_i^{acc}~=~1~-~q_i~+q_i~\omega_i^*~ +~ q_i~n_i~\omega_P~.
%
\label{omega-acc}
}
The HSD results for $\omega_P$ (Fig.~\ref{BBC}, right), $n_i$
(Fig.~\ref{omTeor}, left), $q_i$ (Fig.~\ref{omTeor}, right),
together with  the HSD  nucleon-nucleon values, $\omega_-^*= 3.0$,
$\omega_+^*=2.7$, and $\omega_{ch}^*=5.7$ at $\sqrt{s}=$~200~GeV,
define completely the results for $\omega_i^{acc}$ according to
Eq.~(\ref{omega-acc}). We find a surprisingly  good agreement of
the results given by Eq.~(\ref{omega-acc}) with the PHENIX data
shown in Fig.~\ref{comp}. Note that the centrality
dependence of $\omega_i^{acc}$ stems from the product,
$n_i\cdot \omega_P$, in the last term of the r.h.s. of
Eq.~(\ref{omega-acc}).

In summary, the preliminary PHENIX data \cite{PHENIX} for the
scaled variances of charged hadron multiplicity fluctuations in
Au+Au collisions at $\sqrt{s}=200$~GeV have been analyzed within the
model of independent sources. Assuming that the number of hadron
sources are proportional to the number of nucleon participants,
the HSD transport model was used to calculate the scaled variance
of participant number fluctuations, $\omega_P$, and the number of
$i$'th hadrons per nucleon accepted by the mid-rapidity PHENIX
detector, $q_in_i$, in different Beam-Beam Counter centrality
classes. The HSD model for nucleon-nucleon collisions was also
used to estimate the fluctuations from a single source, $\omega_i^*$.
We find that this model description   is in a good agreement with
the PHENIX data \cite{PHENIX}. In different (5\%) centrality
classes  $\omega_P$ goes down and $q_i n_i$ goes up  with
increasing $\langle N_P\rangle$. This results in  non-monotonic
dependence of $\omega_i^{acc}$ on $\langle N_P\rangle$ as seen in
the PHENIX data.

We conclude that both qualitative and
quantitative features of the centrality dependence of the
fluctuations seen in the present PHENIX data are the consequences
of participant number fluctuations. To avoid a dominance of
the participant number fluctuations one needs to analyze  most
central collisions with a much more rigid ($\le 1\%$) centrality
selection. The statistical model then predicts $\omega_{\pm}<1$
\cite{SM}, whereas the HSD transport model predicts the values of
$\omega_{\pm}$ much larger than unity at $\sqrt{s}=200$~GeV
\cite{KGB}. To allow for a clear distinction between these predictions
it is mandatory to enlarge the acceptance of the
mid-rapidity detector up to about 10\% (see the discussion in
Ref.~\cite{KGB}).

\vspace*{3mm}
{\bf Acknowledgments: }
We like to thank V.V.~Begun, W.~Cassing, M.~Ga\'zdzicki, W.~Greiner,
M.~Hauer, B.~Lungwitz, I.N.~Mishustin and J.T. Mitchell for useful
discussions. One of the author (M.I.G.) is thankful to the Humboldt
Foundation for financial support.


\end{document}